# *Quantum Interference by Vortex Supercurrents*

G. P. PAPARI [1,2,3*] AND V. M. FOMIN [4,5]

[1] *Dipartimento di Fisica, Università di Napoli "Federico II," via Cinthia, I-80126 Napoli, Italy*
[2] *CNR-SPIN, UOS Napoli, via Cinthia, I-80126 Napoli, Italy*
[3] *Istituto Nazionale di Fisica Nucleare (INFN), Naples Unit, via Cinthia, I-80126 Napoli, Italy*
[4] *Institute for Emerging Electronic Technologies, Leibniz IFW Dresden, Helmholtzstraße 20, D-01069 Dresden, Germany*
[5] *Laboratory of Physics and Engineering of Nanomaterials, Department of Theoretical Physics, Moldova State University, str. Alexei Mateevici 60, MD-2009, Chişinău, Republic of Moldova*

Abstract:
We analyze the origin of the parabolic background of magnetoresistance oscillations measured in finite-width superconducting mesoscopic rings with input and output stubs and in patterned films. The transmission model explaining the sinusoidal oscillation of magnetoresistance is extended to address the parabolic background as a function of the magnetic field. Apart from the interference mechanism activated by the ring, pinned superconducting vortices as topological defects introduce a further interference-based distribution of supercurrents that affects, in turn, the voltmeter-sensed quasiparticles. The onset of vortices changes the topology of the superconducting state in a mesoscopic ring in a such a way that the full magnetoresistance dynamics can be interpreted owing to the interference of the constituents of the order parameter induced by both the ring with its doubly-connected topology and the vortex lattice in it.

In a superconducting state, vortices may nucleate to compensate the onset of supercurrents stimulated by magnetic fields [1]. Vortices are quantized excitations composed by supercurrents swirling around single flux lines piercing a "normal" region of the superconductor. These topological defects are stiff and extended in size [2] [3] [4] [5] [6] [7] and affect the path and the phase evolution of coherent Cooper pairs [8].

Measurements of Magnetoresistance Oscillations (MROs) are references to investigate vortex dynamics and their influence on the transport. In these acquisitions, the oscillation of resistance $R$ is usually discussed in terms of the relaxation of supercurrents induced by the onset of successive nucleation of vortices [9] [10] [11] [12] occurring at discrete values of the magnetic field.

The growth of investigations on quantized excitations in superconducting mesoscopic systems is persistent [13] [14] [15] [16], either in shape of simple rings [17] or in combination with semiconducting nanostructures as building blocks for "robust" quantum bits [18] [19] [20].

However, though MROs [21] [22] [23] [24] [25] [26] [27] [28] [29] [30] [31] [32] [33] are recognized as quantum interference effects, the argument deserves further investigations, especially from the phenomenological point of view. In particular, the role of vortices as a source of a topological phase has not been directly connected with the change of measured differential voltage yet.

Quantum interference in both normal and superconducting mesoscopic rings is inherently dependent on geometrical phases gained during the electron transport [34]. The interference shows up through the sinusoidal oscillation of the resistance that −for superconducting samples− is usually superimposed on a parabolic background (PB) [35] [36]. The origin of the PB is attributed to the growth of supercurrents [37] [38] [39], but a dedicated discussion of the way vortex supercurrents affect the onset of the PB is still missing in literature.

In this Letter, we address the latter issue by focusing on the origin of the PB of MROs measured in various types of rings [36] [40] [41] and patterned films [42] [43] [44].

Recently, aiming to derive a model of the MROs in both low and high critical temperature superconductors, we have proposed a reliable approach based on the Ginzburg-Landau theory [45] connecting quantum interference to the onset of the sinusoidal constituent in MROs. Despite previous considerations [40], the interference mechanism affects the dependence of the wave functions of the transmitted particles on spatial coordinates, rather than their density. Both the densities $n_s$ of Cooper pairs (CPs) and thermally activated [46] [47] quasiparticles (QPs), realize the mixture of transmitted particles and depend on $T$ only, while the interference-induced sinusoidal voltage $\Delta V_{int}(T,B) = R(T)\Delta I_{qp}(T,B)$ (see Supporting Information) relies on the variation of the sensed QP charge. Here, the measured zero-bias resistance at $B = 0$ is $R(T)$. We will apply the same model to address the PB by the onset of vortices affecting the topology of the sample as further ring-like objects.

A sketch of a typical mesoscopic ring is represented in Fig. 1, indicating the input and output stubs, through which the voltage across the ring is measured. A magnetoresistance (MR) is acquired in proximity of the critical temperature $T_c$, when the current–voltage characteristic ($IV$) displays a zero-bias resistance due to the thermal activation of phase slips (TAPS) [48] (see Fig. 2(d) in [45]). Usually, to acquire a MR, an AC probe current $I_x$ is injected and a lock-in amplifier measures the voltage across the ring as a function of $B$.

The sinusoidal oscillation in MR has been addressed in terms of the order parameter $\widetilde{\Psi}_r$ of a finite-size ring defined by the inner/outer radius ($r_{i/o}$). $\widetilde{\Psi}_r$ depends on the density of the dia- and paramagnetic supercurrents, distribution of which over the ring's surface oscillates to fulfill the fluxoid quantization (FQ) condition.

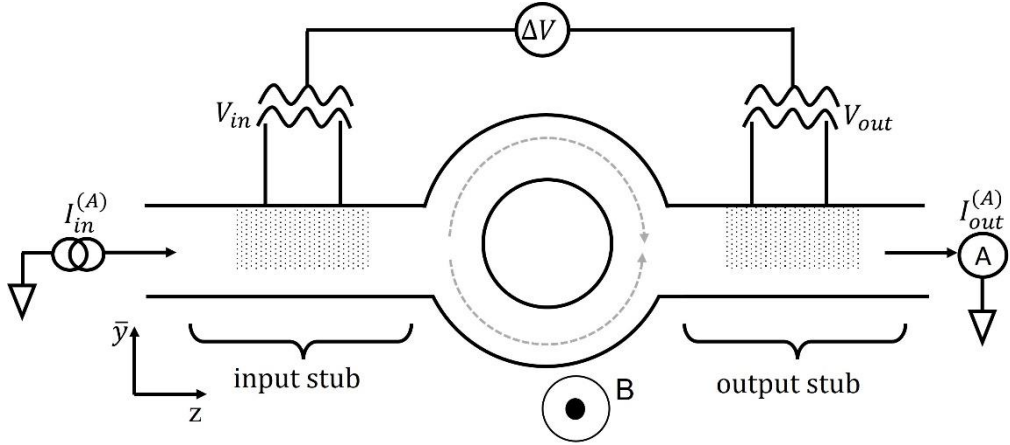

*Figure 1: Sketch of a mesoscopic ring. Input/output voltage and current leads are indicated. The shadowed areas near the voltage leads are the voltage sensitive areas.*

Once the order parameter of the ring is defined, we can evaluate the change in geometrical phase according to the transmission equation [45]

$$\widetilde{\Psi}_{out,ring} = T(\Phi_r)\,\widetilde{\Psi}_{in,ring}, \tag{1}$$

where $T(\Phi_r) = (-1)^n \cos(\pi \frac{\Phi_r}{\Phi_0})$, $n$ is the winding number and $\Phi_r = B\pi r_{avg}^2$ with $r_{avg}$ the average radius of the ring. Information on the change of the transmitted supercurrent density is obtained using the equation

$$\boldsymbol{J}_s = \frac{e\,\hbar}{i\,m^*}\{\widetilde{\Psi}^*\boldsymbol{\nabla}\widetilde{\Psi} - \widetilde{\Psi}\boldsymbol{\nabla}\widetilde{\Psi}^*\} - \frac{2\,e^2|\widetilde{\Psi}|^2}{m^*}\boldsymbol{A}, \tag{2}$$

where $e$ is the electronic charge, $m^*$ is the effective mass of a single electron and $\boldsymbol{A} = \boldsymbol{\nabla} \times \boldsymbol{B}$ is the vector potential of the magnetic field piercing the sample. Substitution of eq. (1) in eq. (2) yields

$$\boldsymbol{J}_{s,out} = T^2(\Phi_r)\,\boldsymbol{J}_{s,in}. \tag{3}$$

The change of the transmitted current density must comply with the conservation of the injected probe current $I_{in} = I_{out}$, so that eq. (3) refers to a change in the spatial density of CPs.

Thermally activated QPs propagate through the domains of the sample where $J_s = 0$ [45] separating the regions of dia- and paramagnetic supercurrents. Thus, supercurrent dynamics also affect the QP spatial density that results in the component of current sensed by voltmeter revealing the interference mechanism.

As previously discussed [45], $B$ stimulates the onset of supercurrents increasing with $n_s$ till a certain critical value $n_{c_1}$ triggers the onset of vortices [2] [49] [50].

Geometry, including the sizes, of the sample strongly affects the vortex supercurrents, especially because of the crowding effects [51] stimulating vortices to nucleate from interconnections between stubs and each of the voltage leads as well as the ring. Yet, confinement and vortex stiffness determine a nonlocal [7] [52] dynamics because local distribution of vortices relies on their motion all over the mesoscopic sample. The onset of dia- and paramagnetic supercurrents [53] along with antivortices [54] make the vortex matter and, consequently, the read out of QP density in each voltage lead very complicated (Fig. 2(a)).

We tackle this problem by defining an average order parameter of the sample realized through the factorization of the swirling supercurrents in the ring $\widetilde{\Psi}_r$ and the vortex supercurrents $\widetilde{\Psi}_V$ (Fig. 2(b)) as shown in the equation

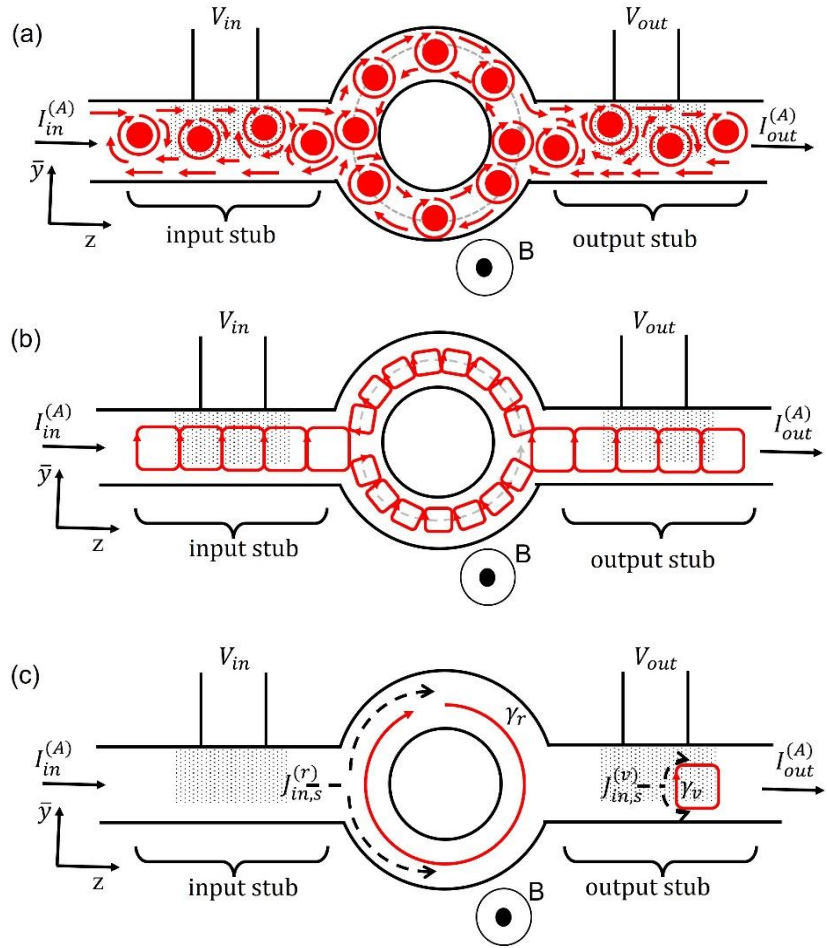

*Figure 2: (a) Scheme of supercurrents flowing in the sample in the presence of vortices. (b) A simplified vortex supercurrent pattern realized by sampling the real distribution by a series of swirling supercurrents at the nanoscale. (c) Reduction of the vortex lattice to a single vortex because of averaging of the vortex supercurrents. Dashed arrows sketch the paths of injected and vortex supercurrents involved in the interference mechanism.*

$$\widetilde{\Psi}_{sample} = \sqrt{n_s}\widetilde{\Psi}_r\widetilde{\Psi}_v\widetilde{\Psi}_\varphi = \sqrt{n_s}\, e^{i\frac{2\pi}{\Phi_0}\oint_{\gamma_r}\Lambda J_{s_r}\cdot dl}\, e^{i\frac{2\pi}{\Phi_0}\sum_{i=1}^{N}\oint_{\gamma_i}\Lambda J_{s_i}\cdot dl_i}\, e^{i\frac{2\pi}{\Phi_0}\left(\oint_{\gamma_r}A\cdot dl + \sum_{i=1}^{N}\oint_{\gamma_i}A\cdot dl_i\right)} \tag{4}$$

where the first factor is the phase contribution owing to the averaged dia- and paramagnetic supercurrents in the ring, the second factor refers to the phase contribution of the vortex supercurrents, whereas the third factor depends on the total flux $\Phi = \oint_{\gamma_r} A\cdot dl + \sum_{i=1}^{N}\oint_{\gamma_i} A\cdot dl_i = \Phi_r + \Phi_v$. The average supercurrent of the ring $J_{s_r}$ is assumed to flow along the circle $\gamma_r = 2\pi r_{avg}$.with the average radius [45] $r_{avg}$. The same approach is applied to the vortex supercurrents swirling along the circular paths $\gamma_i = 2\pi r_i$; $\Lambda = \mu_0 \lambda^2$ where $\lambda$ is the penetration depth and $\Phi_0$ is the flux quantum.

Since the vortex nucleation in mesoscopic structures occurs through the onset of vortex rows [9] [12] [55], the term $\widetilde{\Psi}_v = \prod_{i=1}^{N}\widetilde{\Psi}_i = \exp\{i\frac{2\pi}{\Phi_0}\sum_{i=1}^{N}\oint_{\gamma_i}\Lambda J_i \cdot dl_i\}$ in eq. (4) accounts for the onset of the first vortex row event justifying the PB in MRs. Transport measurements sense the collective coherent dynamics of all vortex supercurrents as if they were generated by a single effective vortex (see Fig. 2(c)), so that

$$\widetilde{\Psi}_v = \exp\{i\frac{2\pi}{\Phi_0}\oint_{\gamma_v}\Lambda J_v \cdot dl_v\}, \tag{5}$$

where $\gamma_v = 2\pi r_v$; if $w$ denotes the sample width, then the condition $2r_v < w$ must be fulfilled. The total phase of $\widetilde{\Psi}_{sample}$ satisfies the FQ

$$\oint_{\gamma_r}(\Lambda J_{s_r} + A)\cdot dl + \oint_{\gamma_v}(\Lambda J_v + A)\cdot dl_v = n\Phi_0, \tag{6}$$

where $n$ is either zero or an integer. According to eqs. (4) and (5), the total transmission equation is

$$\widetilde{\Psi}_{out} = \widetilde{\Psi}_{out,r}\widetilde{\Psi}_{out,v}, \tag{7}$$

so that the total transmission function is

$$T = \cos\left\{\pi\frac{\Phi_r}{\Phi_0}\right\}\cos\left\{\pi\frac{\Phi_v}{\Phi_0}\right\} \tag{8}$$

and $\Phi_v = B\pi r_v^2$. By substituting eq. (7) in eq. (2) and with the support of eq. (8), the modulation of transmitted supercurrent density is obtained in the form

$$J_{s,out} = J_{s,in}^{(r)}\cos^2\left\{\pi\frac{\Phi_r}{\Phi_0}\right\} + J_{s,in}^{(v)}\cos^2\left\{\pi\frac{\Phi_v}{\Phi_0}\right\}, \tag{9}$$

where, as shown in Fig. 2(c), $J_{s,in}^{(r)}$ and $J_{s,in}^{(v)}$ refer to the input supercurrent densities corresponding to the ring with its doubly-connected topology and the effective vortex, respectively. The aforementioned supercurrent densities differ by their spatial distributions as a consequence of the different spatial position of the ring and the effective vortex.

In order to measure the voltage signal $\Delta V_{int}$ related to the quantum interference, the current conservation [45] as sensed by the ammeter (the corresponding quantities are labelled by the index A) can be set

$$I_{in}^{(A)} = I_{out}^{(A)}, \tag{10}$$

and it can be represented as follows

$$I_{in,qp}^{(A)} + I_{in,s}^{(A)} = I_{out,qp}^{(A)} + I_{out,s}^{(A)}, \tag{11}$$

where the subscripts *qp*, *s* refer to the temperature-based partition (see Supporting Information in [45]) into QPs and CPs within the injected current. The growth of $B$ induces vortex supercurrents to swirl either in the input or in the output part of the sample. These contributions $I^{(V)}_{in/out,S_v}$ are not sensed by the ammeter, but can be accounted for within the voltage sensitive areas [45], where the total charge is conserved

$$I^{(V)}_{in,qp} + I^{(V)}_{in,s} + I^{(V)}_{in,S_v} = I^{(V)}_{out,qp} + I^{(V)}_{out,s} + I^{(V)}_{out,S_v}. \tag{12}$$

Magnetic field breaks the symmetry: $I^{(V)}_{in,j} \neq I^{(V)}_{out,j}$ for $j = qp, s, s_v$, and eq. (12) can be written accounting for all increments:

$$I^{(V)}_{out,qp} - I^{(V)}_{in,qp} = I^{(V)}_{in,s} - I^{(V)}_{out,s} + I^{(V)}_{in,S_v} - I^{(V)}_{out,S_v}. \tag{13}$$

According to eqs. (29) and (30) of ref. [45], eq. (13) turns to

$$\frac{\Delta V_{int}(B) + \Delta V_0(T)}{R(T)} = I^{(V)}_{in,s} \sin^2 \pi \frac{\Phi_r}{\Phi_0} + I^{(V)}_{in,S_v} \sin^2 \pi \frac{\Phi_v}{\Phi_0}, \tag{14}$$

where $\Delta V_0(T)$ is the measured voltage for $B = 0$ and $R(T) \equiv \Delta V_0(T)/I_{in}$ is the value of the zero-bias resistance at the temperature of the MR acquisition. Hence, the MR trend accounting for the PB is given by

$$R(T,B) = R(T)\left(1 + \frac{I^{(V)}_{in,s}}{I_{in}} \sin^2 \pi \frac{\Phi_r}{\Phi_0} + \frac{I^{(V)}_{in,S_v}}{I_{in}} \sin^2 \pi \frac{\Phi_v}{\Phi_0}\right), \tag{15}$$

where $R(T,B) = \Delta V_{int}/I_{in}$. It is important, that $I_{in,S_v}$ refers to vortex supercurrents, which, according to the B-T phase diagram, are placed as a function of $B$, and hence of $n_s$, higher than those related to the full supercurrent phase bounded by $B_{c1}(T)$. Thus, the inequality $I_{in} < I^{(V)}_{in,v}$ is physically possible.

The model is validated employing MROs acquired on the "wide" mesoscopic ring reported in [56]. The sample is realized patterning a 30 nm-thick YBa$_2$Cu$_3$O$_{7-\delta}$ (YBCO) film. The ring has the inner and outer radii $r_i \cong 70$ nm and $r_o \cong 190$ nm, respectively. In Fig. 3, black dots and red curves represent the experimental and theoretical data, correspondingly. For three temperatures of acquisition $T = 84.25K, 84.50K, 84.75K$ related to the zero-bias resistances $R(T) = 0.40\Omega, 0.93\Omega, 1.73\Omega$, the best matching parameters are $\frac{I^{(V)}_{in,s}}{I_{in}} = 0.33, 0.15, 0.06$, $\frac{I^{(V)}_{in,S_v}}{I_{in}} = 2.37, 1.17, 0.43$ with an uncertainty of about 5%. The average radii of the vortex and the ring are $r_v = 36$ nm and $r_{avg} = 127$ nm, respectively.

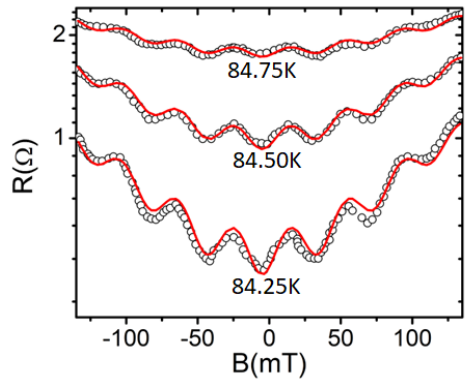

Figure 3: Comparison between the experimental MRs (black circles) of an YBCO mesoscopic ring (after [Ref. [56]]) and the theoretical behavior (red curves) provided by the present model according to eq. (15).

The proposed model is further tested on MROs of superconducting rings with different sizes [45] [57] and nanopatterned films [39] as well (see Supporting Information). The effectiveness of the presented model and its direct dependence on the geometry of multiply connected samples can provide a relevant contribution for solving issues on asymmetric rings presenting oscillation of critical current depending on the direction of injected current [58] [59].

A consistent demonstration of the change in distribution of transmitted QPs must involve a measurement realized through Hall probes at the input and output stubs. Hints on the valuable use of the Hall probes are indirectly provided by the sign-reversal effect [60] [61] measured in planar films. In this regard, CPs, QPs and vortices are considered mutually connected by the interactions, which can provide an edge imbalance charge measurable through the Hall resistance.

In conclusion, we have extended the transmission model developed for the sinusoidal constituent in MROs of superconducting rings to the onset of the PB. We have argued that the onset of vortices and related vortex supercurrents affects the topology of the current flow. Consequently, the full MR dynamics can be addressed through the interference of the constituents of the order parameter induced by both the ring with its doubly-connected topology and the vortex lattice. The PB is inherent to the development of vortex supercurrents, which affect the path of QPs, introducing a further voltage variation that follows a sinusoidal law similarly to the arguments presented for the supercurrents in a ring [45]. Since the vortex size is much smaller than the ring's size, typical narrow-range acquisitions prevent from observation of the background oscillation.

Acknowledgements

This work was supported by the European Cooperation in Science and Technology via COST Action CA21144 (SUPERQUMAP).

# Supplementary information

## *Quantum Interference by Vortex Supercurrents*

G. P. PAPARI [1,2,3*] AND V. M. FOMIN [4,5]

[1] *Dipartimento di Fisica, Università di Napoli "Federico II," via Cinthia, I-80126 Napoli, Italy*
[2] *CNR-SPIN, UOS Napoli, via Cinthia, I-80126 Napoli, Italy*
[3] *Istituto Nazionale di Fisica Nucleare (INFN), Naples Unit, via Cinthia, I-80126 Napoli, Italy*
[4] *Institute for Emerging Electronic Technologies, Leibniz IFW Dresden, Helmholtzstraße 20, D-01069 Dresden, Germany*
[5] *Laboratory of Physics and Engineering of Nanomaterials, Department of Theoretical Physics, Moldova State University, str. Alexei Mateevici 60, MD-2009, Chişinău, Republic of Moldova*

## 1 Classical representation of $\Delta V_{int}$

The measured voltage due to the interference mechanism $\Delta V_{int}$ is dependent on both the resistance at temperature $T$ and the change in the sensed QP current through the expression

$$\Delta V_{int}(T, B) = R(T)\Delta I_{qp}(T, B). \tag{S1}$$

According to eq. (15), $\Delta I_{qp}$ relies on the flux dependent variation of the sensed CPs deriving from the change in the spatial distribution of injected supercurrents. The voltage leads sense different densities of QPs though its total current remains constant. Assuming $R(T)$ as the value of resistance (QP density) measured for $B = 0$, the equivalent classical circuit resorting the interference phenomenon is reported in Fig. S1.

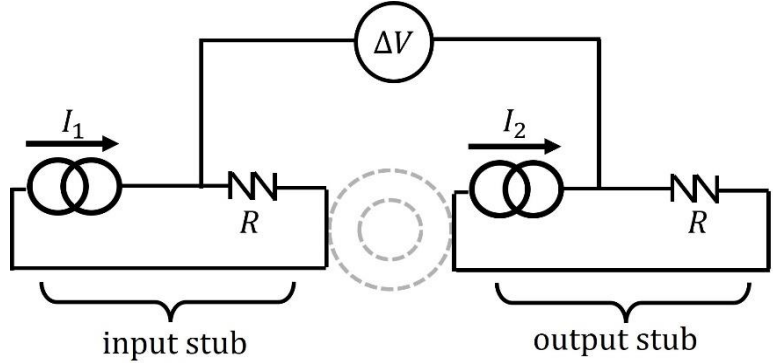

*Figure S2: Equivalent classical circuit interpretating $\Delta V_{int}$.*

The voltage leads measure over two identical circuits presenting the same resistance but fed with different QP currents $I_{1,2}$. From the classical point of view, the voltmeter measurement is a sort of a trans-resistance acquisition because the two stubs are decoupled like the input and output leads of an operational amplifier [1]. According to the scheme of Fig. 1S, we have

$$\Delta V = V_{in} - V_{out} = R(I_1 - I_2). \tag{S2}$$

that has the same structure as eq. (S1) has.

## 2.1 Application of the transmission model to patterned films and stripes

The validation of the transmission model can be extended to other multiply-connected superconducting structures. Here, we focus on three distinct publications presenting MRO on a patterned film [2], on a holed stripe [3] and on a film with an array of pinning centers [4].

Case 1

The patterned film is 26 nm thick and made of $La_{1.84}Sr_{0.16}CuO_4$ (LSCO). The film is patterned as a grid of noninteracting square loops. Each loop has an area of $100 \times 100\ nm^2$ and the wire realizing the web is about $w = 25nm$ wide. The sample presents a full transition temperature at about $T_0 \cong 26K$ (see Fig. 2 of [2]). The MRO employed to validate the model has been acquired at $T = 29.5K$ (see Fig. 6 of [2]). In Fig. 2S(a), a comparison between the experimental (black dots) and the model (red curve) is represented. The experiment shows damped oscillations with a flux periodicity consistent with $B_{c2} \sim 2T$ extracted by eye from Fig. 3 of [2].

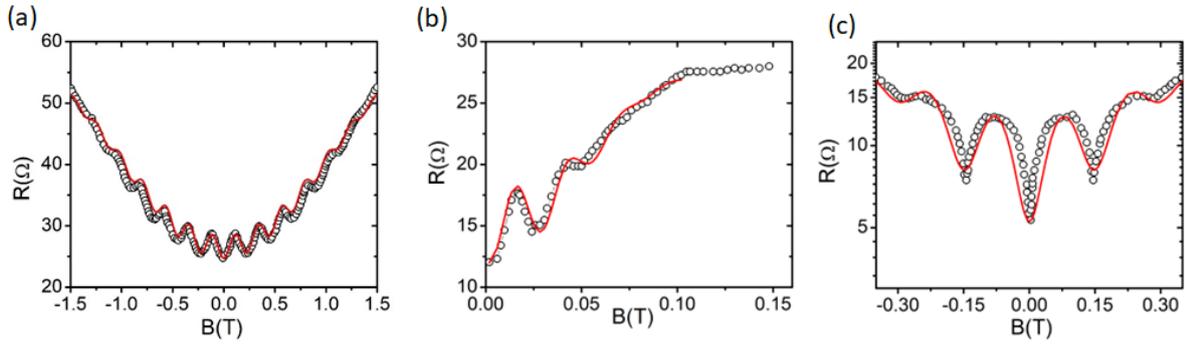

Figure S2: In (a), (b) and (c), the black circles represent the MROs measured in [2],[3] and [4], respectively. Red curves are obtained using our transmission model.

The parabolic background is driven by the first vortex nucleation event that scales like $B_{c1} \sim \frac{\Phi_0}{w^2} = 3.2\ T$ [5], because the magnetic penetration depth is comparable with wires width. Hence, wires are so narrow to make $B_{c1}$ comparable to $B_{c2}$. As a matter of the fact, the experimental trend is similar to the MRs of aluminum rings [6]. The presence of a damping factor modifies eq. (16) in the following form:

$$R_1(T,B) = R(T)\left(1 + \frac{I_{in,s}^{(V)}}{I_{in}}\mathfrak{D}_r\sin^2\pi\frac{\Phi_r}{\Phi_0} + \frac{I_{in,s_v}^{(V)}}{I_{in}}\sin^2\pi\frac{\Phi_v}{\Phi_0}\right), \tag{S3}$$

where $\mathfrak{D}_r = e^{-B^2/B_r^2}$ defines the damping of sinusoidal oscillations as a function of $B_r \sim B_{c1}$ [7]. The fitting parameters are $R(T) = 24.7\ \Omega$, $I_{in,s}^{(V)}/I_{in} = 0.15$, $I_{in,s_v}^{(V)}/I_{in} = 1.25$, $r(\Phi_r) = 53\ nm$, $r(\Phi_v) = 12.5\ nm$ and $B_r = 1.25\ T$.

Case 2

The system investigated in [3] is a Nb slab 100nm thick and 385 nm wide patterned with a series of periodic square holes 120nm in size. The MR curves we are interested in, present a hole period of 385nm and have been extracted from the Fig. 5(a) of [3] (curve made of violet triangles). In Fig. 2S(b), the experimental curves are compared with the model results obtained by us. As distinct from the Case 1, the MR presents the transition up to the normal state reached at about $R_n = 27.9\ \Omega$. Therefore, the vortex supercurrent contribution also presents a damping factor, and the MR curve follows the trend

$$R_2(T,B) = R(T)\left(1 + \frac{I_{in,s}^{(V)}}{I_{in}}\mathfrak{D}_{sq}\sin^2\pi\frac{\Phi_{sq}}{\Phi_0} + \frac{I_{in,s_v}^{(V)}}{I_{in}}\mathfrak{D}_v\sin^2\pi\frac{\Phi_v}{\Phi_0}\right), \tag{S4}$$

where $\mathfrak{D}_{sq} = e^{-B^2/B_{sq}^2}$ and $\mathfrak{D}_v = e^{-B^2/B_v^2}$. The matching of the model with the experimental curves is obtained using $R(T) = 12\ \Omega$, $I_{in,s}^{(V)}/I_{in} = 0.5$, $I_{in,s_v}^{(V)}/I_{in} = 2.3$, $r(\Phi_{sq}) = 151\ nm$, $r(\Phi_v) = 48\ nm$, $B_{sq} = 0.05\ T$ and $B_v = 1.5T$.

Case 3

Large superconducting bridges functionalized with an array of pinning centers display peculiar MRO presenting deep and sharp minima (vortex matching effect), which are placed according to the fields $B_i = \frac{\Phi_0}{a^2}$, where $a$ is the average distance between the pinning centers. These latter can be obtained through two main routes: patterning of antidots [8] or ion irradiation [4]. Pinning centers govern the nucleation of vortices at some specific locations, but the unavoidable presence of interstitial ones promotes the onset of pinned vortices presenting slightly different radii. Hence, the sinusoidal in MR is a function of a series of fluxes $\Phi_{pin,i} = B\pi r_i^2$ with $r_i \in [r_{min}, r_{max}]$ ranging around $a$.

Here, we refer to the measurements conducted over a bridge of YBa$_2$Cu$_3$O$_{7-\delta}$ (YBCO) $50\mu m$ wide and 50nm thick, in which a square array of pinning centers was imprinted through the deposition of a PMMA mask and O$^+$ ion irradiation. The sample is realized with defects having a diameter of 40nm and period of 120nm. Measurement, reported as black dots in Fig. 2S(c), has been extracted from the Fig. 4(c) of [4] referring to the one acquired with the lowest excitation current. The curve has been traced out with the function

$$R_3(T,B) = R(T)\left(1 + \frac{I_{in,s}^{(V)}}{I_{in}}\mathfrak{D}_{pin}\sum_{r_{min}}^{r_{max}} \sin^2\frac{\Phi_{pin}}{\Phi_0} + \frac{I_{in,s_v}^{(V)}}{I_{in}}\sin^2\pi\frac{\Phi_v}{\Phi_0}\right),\quad (S5)$$

where the sum runs over pinned vortices ranging in radius between $r_{min}$ and $r_{max}$, $\mathfrak{D}_{pin} = \exp{-B^2/B_{pin}^2}$ is the damping factor relative to the sinusoidal, and the last term on the right is the interstitial vortices contribution. By employing the parameters $R(T) = 5.4\ \Omega$, $\frac{I_{in,s}^{(V)}}{I_{in}} = 1.3$, $\frac{I_{in,s_v}^{(V)}}{I_{in}} = 2.5$, $r_{min} = 58nm$, $r_{max} = 68nm$, $r_v = 25nm$, $B_{pin} = 0.4T$ we have obtained the red curve in Fig. 2S(c).

## 2.2 Discussion

Damping factors are phenomenological parameters governing the onset of decoherence driven by the magnetic field. If the acquisition range is large, MRO are traced out using damping parameters in both the sinusoidal and background contributions, but the zero-field resistance $R(T)$ is enough to describe the evolution of MRs even if they present a three-fold variation of resistance as shown in the above cases. Do damping factors account for the increase of QP density that grows up to $R_N$? No they do not. In fact, damping factors are different for the geometrical and vortex contribution highlighting that $\mathfrak{D}$'s are not related to the growth of QP density. Damping factors explain the reduction of asymmetry in the distribution of transmitted QPs. The degree of asymmetry, regarding the ring driven interference, relies on the oscillation amplitude of the effective radius that being a function of $J_c/J_s$ [9] lowers if $B$ — and consequently $J_s$ — approaches a critical value. Hence, factors $\mathfrak{D}$ deliver the lowering of $\Delta I_{qp}$ in place of defining the transformation of CPs in QPs. As a matter of fact, $R(T)$ can be assumed constant in eqs. (S3), (S4) and (S5). We argue, in agreement with the phenomenology of Case 2, that the growth of MRs up to $R_N$ can happen keeping the QP density constant.

To discern on vortex or QP constituents of the normal state, more investigations are needed. MRs at high fields must be acquired to realize if the normal state for $B > B_c$ is the same recorded for $T > T_c$. MRs [10] [2] [11] show that the resistive state for $B \sim B_{c2}$ —when the resistance is flat and basically independent on $B$ — is still dependent on $T$ despite the state for $T > T_c$ where the resistance in not dependent on $B$ at all. We speculate that the normal state for $T < T_c$ can be effectively reached only injecting high currents [12] [13] guaranteeing the appropriate heating for the transition to the normal state.

## 3 Enhancement of superconductivity in nanowires

Our model is useful to address the phenomenon of the enhancement of superconductivity observed through the $I_c$ of superconducting nanowires [14] [15] [16] [17]. We argue that the mechanism governing the $I_c$ enhancement is based on the remarkable difference between the nanowire and stub sizes. Fig. S3(a) provides a pictorial representation of the sample under investigation. A nanowire of width $w$ bridges two large stubs, whose transversal dimensions $L_x$, $L_y$ fulfil $w \ll L_x, L_y$. Consequently, the nanowire first critical field $B_{c_n} \sim \Phi_0/w^2$ is quite larger than that for both the left ($L$) and the right ($R$) stubs, which behave as plane films. We focus on the range $B < B_{c_n}$, when the stubs are experiencing the mixed state, whereas the nanowire is fully superconducting. The nanowire decouples the interference phenomena occurring in the two stubs inducing the value of each voltage lead $V^L$ and $V^R$ to depend exclusively on the vortex dynamics of the corresponding stub. This fact is a clear consequence of the standard application of the transmission model that, as usual, starts from the current conservation.

$$I_{in}^L = I_{out}^R, \tag{S6}$$

Which, in terms of QPs and CPs contributions, can be expressed as (in the following the subscript $i$ stands for *in* and $o$ for *out*)

$$I_{i,s}^L - I_{o,s}^R = I_{i,qp}^L - I_{o,qp}^R. \tag{S7}$$

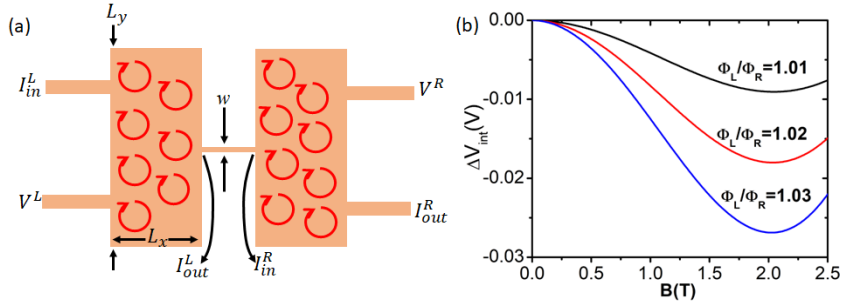

*Figure S3: (a) Pictorial representation of a nanowire sample realized between large stubs. (b) Simulations of $\Delta V_{int}$ as a function of the change in the vortex flux area between the two stubs*

The current conservation also implies that the output current from the left stub is equal to the input one in the right stub, so that

$$I_o^L = I_{o,s}^L + I_{o,qp}^L = I_i^R = I_{i,s}^R + I_{i,qp}^R, \tag{S8}$$

which can be subtracted from eq. (S7) to obtain

$$I_{i,s}^L - I_{o,s}^L + I_{i,s}^R - I_{o,s}^R = I_{o,qp}^L - I_{i,qp}^L + I_{o,qp}^R - I_{i,qp}^R. \tag{S9}$$

Supercurrents and QP contribution can be expressed as [9]

$$I_{o,qp}^L - I_{i,qp}^L = \frac{\Delta V_{int}^L + \Delta V_0}{R(T)} \; ; \quad I_{o,qp}^R - I_{i,qp}^R = \frac{\Delta V_{int}^R + \Delta V_0}{R(T)} \; , \tag{S10}$$

$$I_{i,s}^L - I_{o,s}^L = I_{i,s}^L \sin^2 \pi \frac{\Phi_L}{\Phi_0}; \quad I_{i,s}^R - I_{o,s}^R = I_{i,s}^R \sin^2 \pi \frac{\Phi_R}{\Phi_0} \tag{S11}$$

allowing to rewrite eq. (S9) in this way:

$$I_{i,s}^L \sin^2 \pi \frac{\Phi_L}{\Phi_0} + I_{i,s}^R \sin^2 \pi \frac{\Phi_R}{\Phi_0} = \frac{\Delta V_{int}^L + \Delta V_0^L}{R(T)} + \frac{\Delta V_{int}^R + \Delta V_0^R}{R(T)}, \tag{S12}$$

where $\Delta V_0^{L,R}$ represents the voltage of each stub for $B = 0$ with respect to the ground. Eq. (S12) means that the nanowire decouples the two vortex dynamics so that the following equations hold true

$$\begin{cases} \frac{\Delta V_{int}^L + \Delta V_0^L}{R(T)} = I_{i,s}^L \sin^2 \pi \frac{\Phi_L}{\Phi_0}, \\ \frac{\Delta V_{int}^R + \Delta V_0^R}{R(T)} = I_{i,s}^R \sin^2 \pi \frac{\Phi_R}{\Phi_0}. \end{cases} \quad (S13)$$

Eqs. (S13) enable one to evaluate the measured voltage drop across the device $\Delta V_{int} = \Delta V_{int}^R - \Delta V_{int}^L$ to be

$$\Delta V_{int}(T,B) = R(T) \left( I_{i,s}^R \sin^2 \pi \frac{\Phi_R}{\Phi_0} - I_{i,s}^L \sin^2 \pi \frac{\Phi_L}{\Phi_0} \right) + const, \quad (S14)$$

where $const = \Delta V_0^R - \Delta V_0^L$. Eq (S14) drives the variation of the voltage range, in which the $I_c$ is extracted. As discussed in the article, the lowering of $\Delta V_{int}$ versus $B$ implies the enhancement of the critical current that is the amount of injected current $I_c(V)$ relative to a voltage drop $V \leq V_{th}$. In Fig. S3(b), a simulation of $\Delta V_{int}$ is reported according to a slight change in the effective vortex flux area between the two stubs. Setting a mismatch between $\Phi_L$ and $\Phi_R$ of about a few percent, we obtain a result similar to what was measured in [14], in which $I_c$ grew up to about 2T.